\documentclass[12pt]{article}
\usepackage{graphicx,amsmath,amssymb,color,bm}
\begin{document}

\centerline {\Large\textbf {Optical Properties of Monolayer Tinene in Electric Fields
}}


\centerline{Rong-Bin Chen$^{1}$, Chin-Wei Chiu$^{2}$, and Szu-Chao Chen$^{3}$}

\centerline{$^{1}$Center of General Studies, National Kaohsiung Marine University, Kaohsiung 811, Taiwan}
\centerline{$^{2}$Department of Physics, National Kaohsiung Normal University, Kaohsiung 824 , Taiwan}
\centerline{$^{3}$Department of Physics, National Cheng Kung University, Tainan 701, Taiwan }

\begin{abstract}

Monolayer tinene presents rich absorption spectra in electric fields. There are three kinds of special structures, namely shoulders, logarithmically
symmetric peaks and asymmetric peaks in the square-root form, corresponding to the optical excitations of the extreme points, saddle points and
constant-energy loops. With the increasing field strength, two splitting shoulder structures, which are dominated by the parabolic bands of ${5p_z}$ orbitals,
come to exist because of the spin-split energy bands. The frequency of threshold shoulder declines to zero and then linearly grows. The third shoulder at ${0.75 \sim 0.85}$ eV
mainly comes from (${5p_x,5p_y}$) orbitals. The former and the latter orbitals, respectively, create the saddle-point symmetric peaks near the M point,
while they hybridize with one another to generate the loop-related asymmetric peaks. Tinene quite differs from graphene, silicene, and germanene.
The special relationship among the multi-orbital chemical bondings,  spin-orbital couplings and Coulomb potentials accounts for the feature-rich optical properties.

\vskip0.6 truecm


\end{abstract}

\newpage

\bigskip

\centerline {\textbf {I. INTRODUCTION}}%

The layered condensed-matter systems, with the nano-scaled thickness, have attracted a lot of theoretical \cite{YZheng, VPGusynin, CCLiu, YXu, MEzawa} and experimental \cite{BRadisavljevic, KHao, KSNovoselov, YZhang, LTao, MEDavila, FFZhu} researches, mainly owing to the unusual geometric symmetry and the rich intrinsic interactions. Specifically, the 2D group-IV systems have the hexagonal symmetry, the unique orbital hybridizations, and the significant spin-orbital couplings (SOCs). They are the ideal 2D materials for studying the diverse physical, chemical and material phenomena \cite{YZheng, VPGusynin, CCLiu} and are expected to have high potentials in the near-future technological applications \cite{LTao, ZSun, FSchwierz}. Graphene could be produced by various experimental methods \cite{KSNovoselov, YZhang, KSNovoselov2, MMurakami, MZhang, YShibayama, AMAffoune} since the first discovery in 2004. Silicene and germanene have been fabricated on metallic substrates \cite{LTao, MEDavila, PVogt, BAufray, LFLi, MDerivaz}. Recently, tinene is successfully synthesized on a substrate of bismuth telluride \cite{FFZhu}. In addition, monolayer Pb system is absent in the experimental growth up to now. Graphene has a very strong $sp^2$ bonding on a planar structure, while silicene, germanene and tinene possess the buckled structure with a mixed $sp^2$-$sp^3$ hybridization. The spin-orbital coupling is almost negligible in the former, but very important for the latter. Among three buckled systems, tinene exhibits the strongest $sp^3$ bonding and SOC. These two critical interactions can dramatically change the electronic energy spectra and thus create the rich optical properties. For monolayer tinene, absorption spectra, which arise from the cooperation of the orbital- and spin-dependent interactions and the external electric field (${\bf F\,}$${=F\hat z}$), will be explored in detail. A comparison with other group-IV systems is also made.

It is well known that the low-energy essential properties of monolayer graphene are dominated the ${2p_z}$-orbital hybridization and the hexagonal honeycomb symmetry. This system is a zero-gap semiconductor, with a vanishing density of states (DOS) at the Fermi level (${E_F=0}$), since there exists an isotropic Dirac cone formed by the linearly intersecting valence and conduction bands. Silicene and germanene possess the significant SOCs much stronger than that in graphene. Such SOCs can separate the Dirac-cone structures built from the dominating ${3p_z}$ or ${4p_z}$ orbitals, so that the intrinsic systems are direct-gap semiconductors with ${E_g\,\sim 5}$ and 45 meVs for Si $\&$ Ge, respectively \cite{CCLiu, CCLiu2}. As to tinene, the theoretical calculations predict that the low-lying energy bands arising from the (${5p_x,5p_y,5p_z}$) orbitals appear simultaneously, directly reflecting the tight cooperation between the serious $sp^3$ hybridizations and the strong SOCs \cite{SCChen}. A small indirect gap of ${E_g\,\sim}$54 meV is mainly determined by the separated Dirac-cone structure and the ($5p_x$,$5p_y$)-dominated energy bands.
 Moreover, an external electric field can be applied to modulate the essential properties of the buckled systems, such as, the obvious changes in band gap, energy dispersions and state degeneracy \cite{MEzawa, CJTabert}.

In this work, the tight-binding model \cite{YHLai, CYLin} and the gradient approximation \cite{CLLu, MFLin} are, respectively, utilized to evaluate electronic and optical properties of monolayer tinene. The $sp^3$ chemical bonding, SOCs and electric field are included in the calculations. The  orbital- and $F_z$-dependent absorption spectra are thoroughly investigated. This study shows that the feature-rich electronic structures are responsible for the diverse optical spectra. There are three kinds of critical points in the energy-wave-vector space, including the extreme and the critical points, and the  constant-energy loops centered at certain extreme points. They, respectively, lead to the shoulder structure, the logarithmic-form symmetric peaks and the square-root asymmetric peaks in absorption spectra. Furthermore, the threshold frequency and the initial structures are easily tuned by an electric field. Tinene sharply contrasts with graphene, silicene and germanene in the main features of optical properties, e.g., the form, number, frequency and intensity of the low- and middle-frequency absorption structures. The predicted results could be verified by the optical spectroscopy \cite{LReining}.

\bigskip
\bigskip
\centerline {\textbf {II. METHODS}}%
\bigskip
\bigskip

Monolayer tinene has a buckled structure, in which the A $\&$ B equivalent sublattices are located at two parallel planes with a separation of $2l$ ($l$=0.42 $\AA$; Fig. 1(a)) . The lattice constant in the same sublattice is 4.7 $\AA$, and the angle between the Sn-Sn bond and the z-axis is 107.1$^\circ$.
There are two Sn atoms in a primitive unit cell. Each atom contributes
(${5s,5p_x,5p_y,5p_z}$) orbitals to electronic structures; furthermore, the spin-up and spin-down configurations dominate the on-site SOCs.
The Hamiltonian built from the orbital- and spin-dependent tight-binding function is a ${16\times\,16}$ Hermitian matrix. In the presence of a uniform perpendicular electric field, the tight-binding Hamiltonian, with the nearest-neighbor interactions, is expressed as

{\small
\begin{align}
H& =\underset{\langle i\rangle ,o,\text{ }s}{\sum }%
E_{o}C_{ios}^{+}C_{ios}^{{}}+\underset{\langle i,j\rangle ,o,o^{\prime },%
\text{ }s}{\sum \gamma _{oo^{\prime }}^{\overrightarrow{R}%
_{ij}}C_{ios}^{+}C_{jo^{\prime }s}^{{}}+}\underset{\langle i\rangle
,p_{\alpha },\text{ }p_{\beta },s,s^{\prime }}{\sum }\frac{\lambda _{soc}}{2}%
C_{ip_{\alpha }s}^{+}C_{ip_{\beta }s^{\prime }}^{{}}(-i\epsilon _{\alpha
\beta \gamma }\sigma _{ss^{\prime }}^{\gamma })  \notag \\
& +l\underset{\langle i\rangle ,o,\text{ }s}{\sum }\mu
_{i}FC_{ios}^{+}C_{ios}^{{}}\ ,
\end{align}%
}
where $i(j)$, $o$($o^\prime$), and $s$($s^\prime$) represent the lattice site,
atomic orbital, and spin, respectively. The site energy of ${5s}$ orbital is -6.23 eV, as measured from that of the $5p$ orbitals. $E_{o}$ and $\gamma _{oo^{\prime }}^{%
\overrightarrow{R}_{ij}}$ are site energy and hopping integral,
respectively. The latter depends on the type of atomic orbitals with the
 nearest-neighbor vector $\overrightarrow{R}_{ij}$, in which the Slater-Koster hopping interactions are
V$_{ss\sigma}$ =$-$2.6245 eV, V$_{sp\sigma}$=2.6504 eV,
V$_{pp\sigma}$ =1.4926 eV, and  V$_{pp\pi}$ =$-$0.7877 eV \cite{CCLiu2, JCSlater}.
 The third term of $\lambda _{soc}$= 0.8 eV is, the effective SOC on the same atom. $\alpha $, $%
\beta $, and $\gamma $ denote the x, y, and z components, and $\sigma $ is
the Pauli matrix. The last term is the $F$-induced Coulomb potential energy with $\mu _{i}=+1(-1)$ for the A(B) site.

\bigskip {When monolayer tinene exists in an electromagnetic  field, electrons are excited from the occupied states to the unoccupied ones under the vertical transitions. The initial and final states  have the same wave vector (${\Delta\,k_x\,=0,\Delta\,k_y=0}$) as a result of the almost vanishing photon momentum. Based on the Fermi's golden rule, the optical absorption
is given by}

{\small
\begin{align}
{\ }A(\omega )& \propto \underset{h,h^{\prime},n,n^{\prime }}{\sum }\int_{1^{st}BZ}%
\frac{dk_{x}dk_{y}}{(2\pi )^{2}}\left\vert \left\langle \Psi _{n^{\prime
}}^{h^{\prime}}(k_{x},k_{y})\right\vert \frac{\widehat{E}\cdot \mathbf{P}}{m_{e}}%
\left\vert \Psi _{n}^{h}(k_{x},k_{y})\right\rangle \right\vert ^{2}  \notag
\\
& \times Im[\frac{f(E_{n^{\prime
}}^{h^{\prime}}(k_{x},k_{y}))-f(E_{n}^{h}(k_{x},k_{y}))}{E_{n^{\prime
}}^{h^{\prime}}(k_{x},k_{y})-E_{n}^{h}(k_{x},k_{y})-(\omega \mathbf{+}%
i\gamma )}],
\end{align}
}
where ${f(E_{n}^{h}(k_{x},k_{y}))}$ is the
Fermi-Dirac distribution function, $\widehat{E}$ the
unit vector of an electric polarization, and $\gamma$ (=10 meV) the broadening parameter. Energy band (${E^h_n(k_x,k_y)}$) and wave function (${\Psi^h_n\,(k_x,k_y)}$) are obtained from diagonalizing the Hamiltonian in Eq. (1).
$h$ represents the valence or conduction band,  and $n$ corresponds to the $n$th band measured from the Fermi level. An electric polarization of $\widehat{E}$ $\parallel
\widehat{x}$ is chosen for a model study at zero temperature, since the direction-dependence of absorption spectrum is negligible. The velocity matrix element in Eq. (2) is approximated by the gradient of Hamiltonian versus $k_x$. Similar  approximations have been successfully applied to comprehend optical spectra of carbon nanotubes \cite{MFLin} and  few-layer graphenes \cite{CLLu}.

\bigskip
\bigskip
\centerline {\textbf {III. RESULTS AND DISCUSSION}}%
\bigskip
\bigskip

\bigskip  Monolayer tinene exhibits three pairs of valence and conduction bands in the range of $-$3 eV${\le\,E^{c,v}\le\,3}$ eV, as clearly shown in Fig. 1(b). They are mainly determined by the ${(5p_x,5p_y,5p_z)}$ orbitals (Fig. 1(c)); therefore, the $5s$ orbitals make  important contributions only at deeper or higher energies. Each band is doubly  degenerate in the spin degree of freedom; that is, the up- and down-dominated electronic states are degenerate to each other. The low-lying energy bands near the K/K$^\prime$ and $\Gamma$ points appear simultaneously. Without the SOC, the isotropic Dirac-cone structure due to the ${5p_z}$ orbitals is gapless at the corners of the first Brillouin zone (black curve in Fig. 1(b)). The first and second pairs of energy bands near the $\Gamma$ point, with double degeneracy, mainly come from the (${5p_x,5p_y}$) orbitals. Specifically, the SOC creates the slightly deformed Dirac cones with an energy spacing of ${\sim\,0.12}$ eV and the splitting/anti-crossing of the first and second valence (conduction) bands (green curves). The lowest unoccupied state and the highest occupied state are, respectively, situated at the K/K$^\prime$ and $\Gamma$ points, i.e., there exists an indirect gap of ${\sim\,54}$ meV. Electronic structures have three kinds of energy bands, namely, linear, parabolic and partial flat energy dispersions (centered at the K or $\Gamma$ point). Apparently they reveal the critical
points in the energy-wave-vector space (Fig. 2). The extreme (minima $\&$ maxima) and  saddle points, respectively, correspond to the (K/K$^\prime$,$\Gamma$) and M ones (triangles and squares in Fig. 1(b); some of them in Figs. 2(a)-2(c)). It is also noticed that the unusual constant-energy loops occur in between the $\Gamma$ and M (K/K$^\prime$) points (circles in Fig. 1(b)). Those of the first and second pairs of energy bands are due to the cooperation of the $sp^3$ bonding and SOC (black and green curves in Fig. 1(b)), e.g., that of the first valence band in Fig. 2(b). The feature-rich energy dispersions and critical points will dominate the special structures in DOSs and absorption spectra. They could be directly examined by the angle-resolved photo-emission spectroscopy (ARPES) \cite{TOhta, DASiegel, KSKim, TOhta2}.

{\small DOS of monolayer tinene, defined as

\bigskip {\small
\begin{equation}
{\ }D(E)=\underset{n,h=c,v}{\sum }\int_{1^{st}BZ}\frac{dk_{x}dk_{y}}{%
(2\pi )^{2}}\frac{\gamma }{\pi }\frac{1}{[E
-E_{n}^{h}(k_{x},k_{y}]^{2}+\gamma ^{2}},
\end{equation}
presents a lot of special structures in Fig. 1(d). The shoulder structures (triangles), the logarithmic-form symmetric peaks (squares), the square-root peaks (circles) and the delta-function-like peaks (crosses), respectively, originate from the extreme points, the saddle points, the constant-energy loops with infinite extreme points (regarded as the 1D parabolic bands), and the partially flat energy dispersions. The first and second (third) structures might be merged together due to the close energies. The above-mentioned special structures could be further divided into the orbital-created ones. The $5p_z$-dependent features cover a pair of shoulder structures centered at $E=0$ (K in the inset), two symmetric peaks at ${E=\pm\,0.9}$ eV, and a symmetric peak  at ${E=-2.7}$ eV. The ${(5p_x,5p_y,5p_z)}$ orbitals co-dominate all the
asymmetric peaks. The other shoulder structures and symmetric peaks are closely related to the ${(5p_x,5p_y)}$ orbitals. The STS measurements, in which the tunneling conductance (dI/dV) is proportional to DOS, can serve as efficient methods to examine the special structures in DOS. They have been successfully utilized to verify the diverse electronic properties in few-layer graphenes \cite{PLauffer, DPierucci, MYankowitz, GLi}, graphene nanoribbons \cite{LTapaszto, CTao}, and carbon nanotubes \cite{JWWilder, TWOdom, TWOdom2}.  The experimental verifications on the rich structures could provide the useful information about the orbital- and SOC-dominated energy bands in tinene.

Absorption spectra strongly depend on joint density of states (JDOS) and the velocity matrix elements. The former represents all the available excitation channels under the vertical transitions. JDOS, as clearly indicated in Fig. 3(a), has a lot of special structures due to the initial and final critical points. ${A(\omega\,)}$, which directly reflects the main features of energy dispersions, exhibits two low-frequency shoulder structures at ${\omega\,=0.12}$ $\&$ 0.75 eVs (red curve in Fig. 3(b)). The threshold shoulder originates from the excitation between two separated Dirac points (Fig. 1(b); K$_{1\rightarrow\,1}$) and the second one from the first pair of energy bands near the $\Gamma$ point (${\Gamma_{1\rightarrow\,1}}$). They are dominated by the ${5p_z}$ and (${5p_x,5p_y}$) orbitals, respectively. Also, the first pair of valence and conduction bands can create a logarithmic-form  peak and a square-root peak at ${\omega\,=1.75}$ $\&$ 2.94 eVs (Fig. 3(c)), respectively, corresponding to the M saddle points and constant-energy
loops (M$_{1\rightarrow\,1}$ and C$_{1\rightarrow\,1}$}). The symmetric one associated with the ${5p_z}$ orbitals has absorption frequency about double that of ${V_{pp\pi}}$, as observed in graphene, silicene and germanene [Fig. 5]. The asymmetric one due to the (${5p_x,5p_y,5p_z}$) orbitals is determined by the SOC and ${sp^3}$ bonding.

The spectral structures of optical excitations, which are closely related to the second and third pairs of energy bands, occur at the middle-frequency range of 2 eV${\le\,\omega\le}$5 eV (Fig. 3(c)). Part of them arise from the intra-pair optical excitations, including the extreme and saddle points (${\Gamma_{2\rightarrow\,2}}$ and M$_{2\rightarrow\,2}$); the constant-energy loops in the second
pair (C$_{2\rightarrow\,2}$). There exist the significant inter-pair optical excitations, such as the structures                           due to $\Gamma_{3\rightarrow\,2}$, M$_{1\rightarrow\,2}$, C$_{1\rightarrow\,2}$, C$_{2\rightarrow\,1}$, and  C$_{2\rightarrow\,3}$. In addition, few structures associated with the deeper- or higher-energy bands are not assigned the specific excitation channels. The feature-rich absorption spectra clearly illustrate  the strong cooperation between the ${sp^3}$ bonding and SOC.

The electric field can dramatically change band structure and thus absorption spectrum, especially for the low-energy essential properties. $F$ destroys the mirror symmetry about the $z=0$ plane, leading to the spin-split energy bands (Fig. 4(a)). With the gradual increase of $F$, the spin up-dominated valence and conduction bands
near the Dirac points approach to $E_F$ (solid curves). Energy gap is vanishing at the critical field $F_c=0.176$ V${/\AA}$, in which the linear Dirac-cone structure is recovered. Moreover, $E_g$ is opened under the further increase of $F$. However, the spin down-dominated energy bands deviate from $E_F$ monotonously (dashed curves). As a result, the optical threshold shoulder becomes two splitting ones, as clearly indicated in Fig. 3(b) at various $F$s. The absorption frequency of the first structure declines, reaches zero, and then grows in the increment of $F$ (Fig. 4(b)). That of the second one increases smoothly. The similar results could be observed in germanene and silicene (Figs. 4(c) and 4(d)). On the other hand, energy bands near the $\Gamma$ point remain double degeneracy and makes main contributions to the third shoulder  with the monotonous $F$-dependence (the third absorption frequency in Fig. 4(b)). In short, the cooperation of the Coulomb potential and SOC results in the spin-split electronic states near the K and K$^\prime$ valleys, but not the $\Gamma$ valley.

Tinene is in sharp contrast to other group-IV systems in band structures and optical properties, as indicated from Figs. 1 and 3-5. Germanene, silicene and graphene do not have the low-lying energy bands near the $\Gamma$ point, so that they exhibit the direct or vanishing energy gaps (Figs. 5(a)-5(c)) and the lower threshold absorption frequencies (Figs. 4(c) and 4(d)). The parabolic bands centered at the $\Gamma$ point possess the deeper or higher state energies; therefore, the second shoulder structure cannot be observed in the range of ${\omega\,<}$ 1 eV. The constant-energy loop is absent in the first valence band, leading to the disappearance of asymmetric peaks in absorption spectra. The number of critical points within ${|E^{c,v}|\le\,3}$ eV is less than that of tinene, and so do the spectral structures. Among these four systems, graphene has the highest absorption frequency in the M${_{1\rightarrow\,1}}$ excitations (the $\pi$-electronic excitations), owing to the strongest hopping interaction of the $p_z$ orbitals
($V_{pp\pi}$). The strengths of orbital hybridizations and SOC can account for the above-mentioned critical differences. On the other hand, tinene, germanene and silicene exhibit the narrow gaps and the $F$-dependent spltting effects on the low-energy essential properties.

Optical spectroscopies \cite{ZQLi, LMZhang, KFMak, YZhang2, CHLui} have been utilized to identify the rich excitation properties in the low-dimensional carbon-related systems, such as,  few-layer graphenes \cite{ZQLi, LMZhang, KFMak, YZhang2, CHLui} and carbon nanotubes \cite{KLiu, JCBlancon}. Such systems possess the ${\sim5-6}$-eV strong $\pi$ peak corresponding to the optical excitations of saddle points. This peak, which arises from the $\pi$ bondings of $p_z$ orbitals, is predicted to exist in other group-IV 2D systems (Figs. 3 and 5). The AB- and ABC-stacked graphenes are different from of $p_z$ each other  in the absorption frequencies, spectral structures and  $F$-induced excitation spectra \cite{ZQLi, LMZhang, KFMak, YZhang2, CHLui}. Moreover, carbon nanotubes exhibit the strong dependence of
asymmetric absorption peaks on radius and chirality \cite{KLiu, JCBlancon}. The unique optical properties of tinene are worthy of detailed experimental examinations, including
shoulders, symmetric peaks, asymmetric peaks, and $F$-dependent excitation frequencies and absorption structures. These directly reflect the unusual critical points and the coupling effect of $sp^3$ bondings and SOC. The optical experiments, accompanied with the ARPES and STS measurements, are very useful in fully understanding the critical orbital hybridizations and SOC and thus the important differences among group-IV 2D systems.

\bigskip
\bigskip
\centerline {\textbf {IV. Concluding Remarks}}%
\bigskip
\bigskip

By the tight-binding model and gradient approximation, we have calculated for monolayer tinene the electronic structure, DOS and absorption spectrum. The close relationship among the $sp^3$ bonding, SOC and electric field is responsible for the diverse essential properties. Tinene quite differs from germaene, silicene and graphene in band structures and optical spectra, mainly owing to the strength of orbital hybridizations and SOC. The predicted results could be verified by ARPES, STS, and optical spectroscopy.

The main features of energy bands lie in the critical points and the orbital-,
spin- and $F$-dependence. The form, the extreme and saddle points $\&$ the constant-energy loops, have the unusual DOS and spectral structures. They exhibit the shoulders, and the symmetric and asymmetric peaks in DOS, respectively.
The K/K$^\prime$ valleys and the first pair of saddle points mainly come from the $5p_z$ orbitals. The ${(5p_x,5p_y,5p_z)}$ co-dominate the constant-energy loops. The other critical points are closely related to the ${(5p_x,5p_y)}$ orbitals. The spin up- and down-dominated energy bands in the deformed Dirac cones are split by an external electric field, leading to a non-monotonous dependence of $E_g$ on $F$. All the critical points create a lot of special structures in absorption spectra under the vertical transitions of the intra- and inter-pair of energy bands. Specifically, the structures due to the first pair of energy bands cover the first and second shoulders, a symmetric peak and a asymmetric peak, corresponding to the (K,$\Gamma$,M) points and constant-energy loops, respectively. The $\Gamma$-induced low-frequency shoulder and the asymmetric peak are absent in other group-IV systems. The first shoulder becomes two splitting ones in the presence of $F$. Moreover, there exist exist many structures above the middle frequency, being associated with three pairs of energy bands. In addition, germanene and silicene  exhibit the $F$-dependent splitting behavior, and all the systems  have the prominent  $\pi$-electronic absorption peaks. The experimental measurements on energy bands, DOS and optical spectra could identify the orbital hybridizations, the strength of SOC and the effects of electric field in 2D  group-IV systems.

\bigskip

\bigskip

\centerline {\textbf {ACKNOWLEDGMENT}}%

\bigskip

\bigskip

\noindent \textit{Acknowledgments.} This work was supported by the MOST of Taiwan, under Grant No. MOST 105-2112-M-006-007-MY3.


\newpage

\begin{figure}
\centering
\includegraphics[width=1\textwidth]{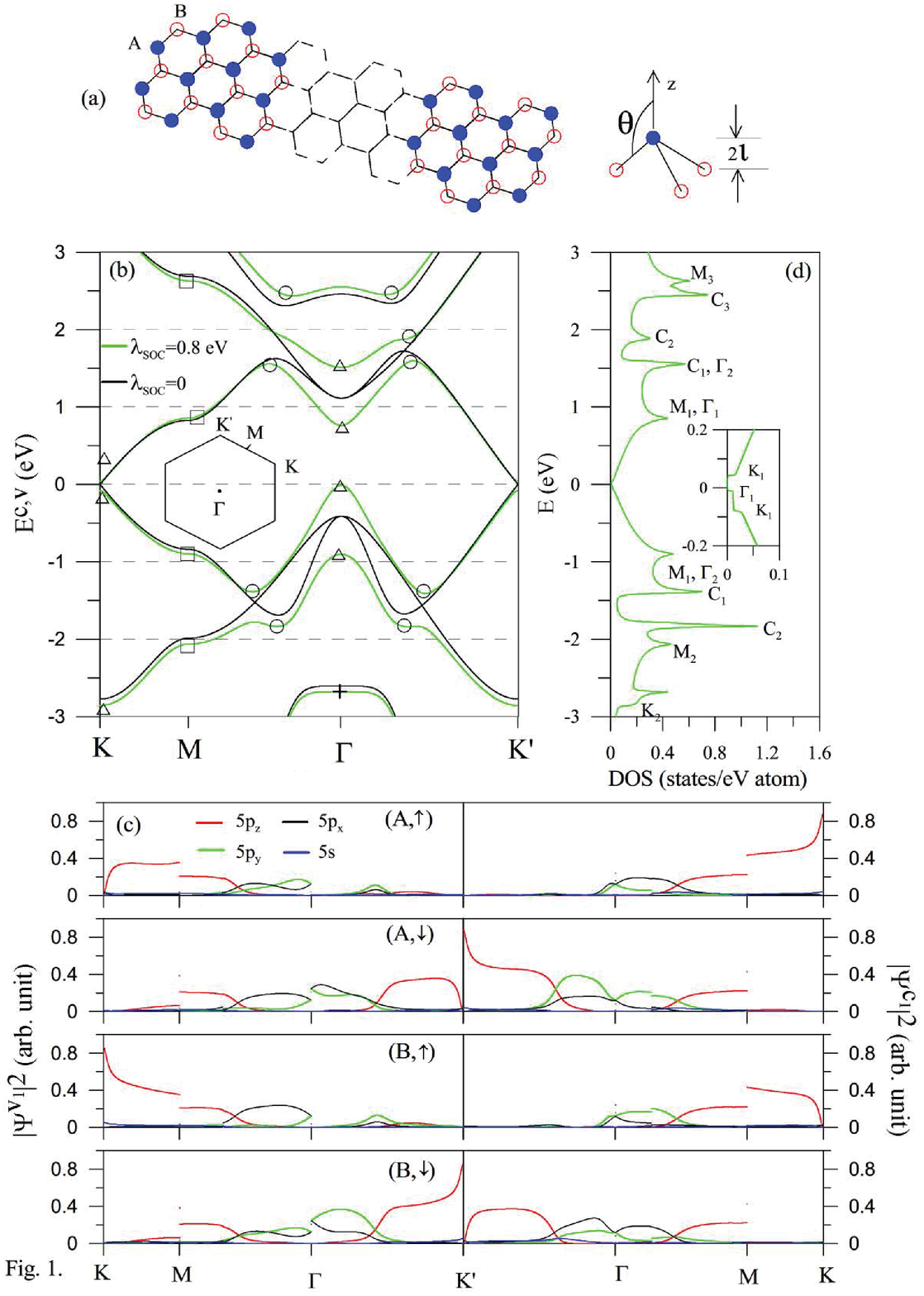}
\caption{For monolayer tinene: (a) $xy$-plane geometric structure $\&$ ${sp^3}$ bonding, (b) three pairs of energy bands along the high symmetry points $\&$ the first Brillouin zone, (c) orbital-dependent wave-function probabilities for the first pair of energy bands; (d) density of states $\&$ the low-energy result in the inset. The subscripts in (c) are associated with the $n$-th energy band measured from the Fermi level.}
\label{FIG:1}
\end{figure}

\begin{figure}
\centering
\includegraphics[width=1\textwidth]{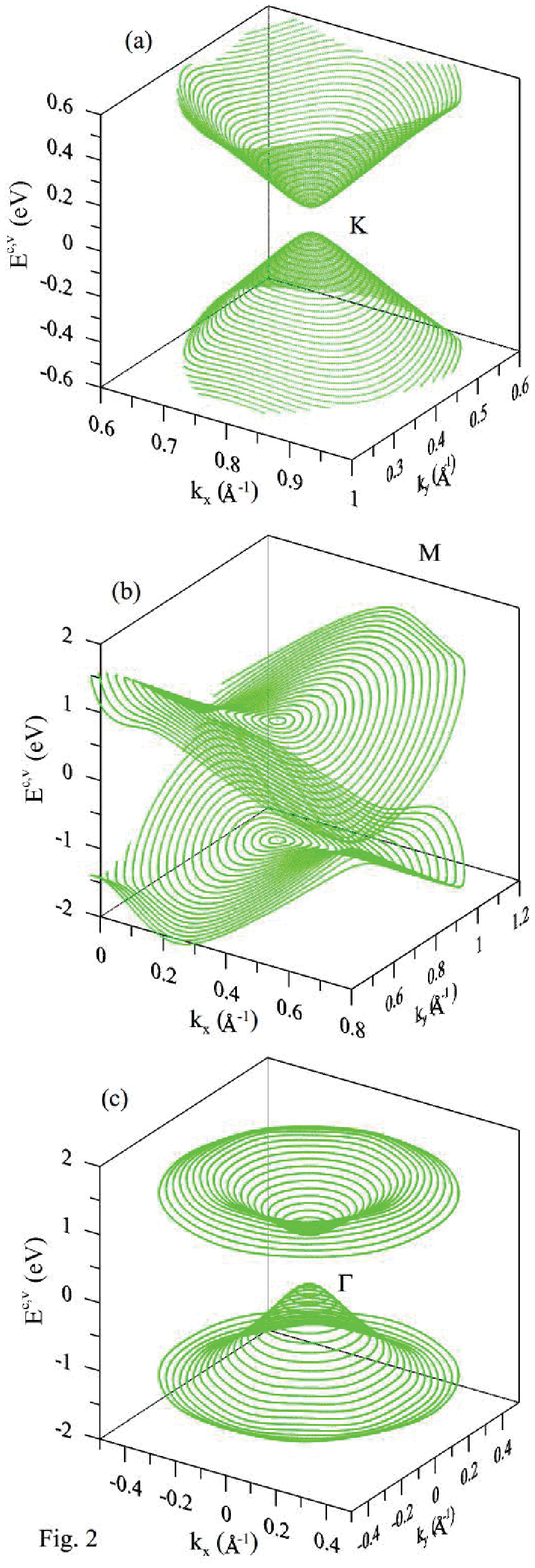}
\caption{The low-lying electronic structures in the energy-wave-vector space near the (a) K, (b) M and (c) $\Gamma$ points.}
\label{FIG:2}
\end{figure}

\begin{figure}
\centering
\includegraphics[width=0.85\textwidth]{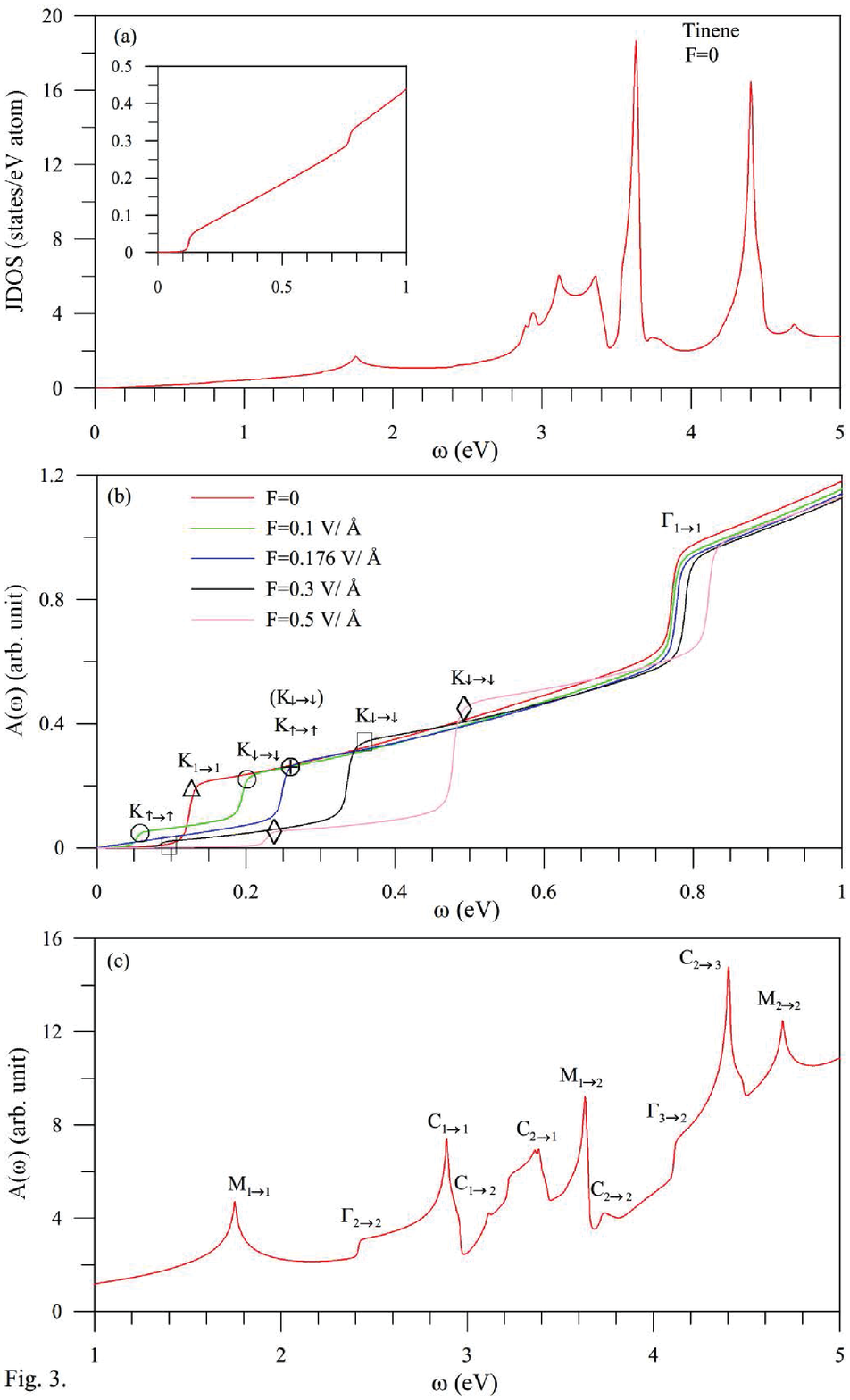}
\caption{Optical properties of tinene: (a) joint density of states; the (b) low- and (c) middle-frequency absorption spectra. Inset in (a) is the low-$\omega$ JDOS. Also shown in (b) is the  $F$-induced spectra.}
\label{FIG:3}
\end{figure}

\begin{figure}
\centering
\includegraphics[width=0.9\textwidth]{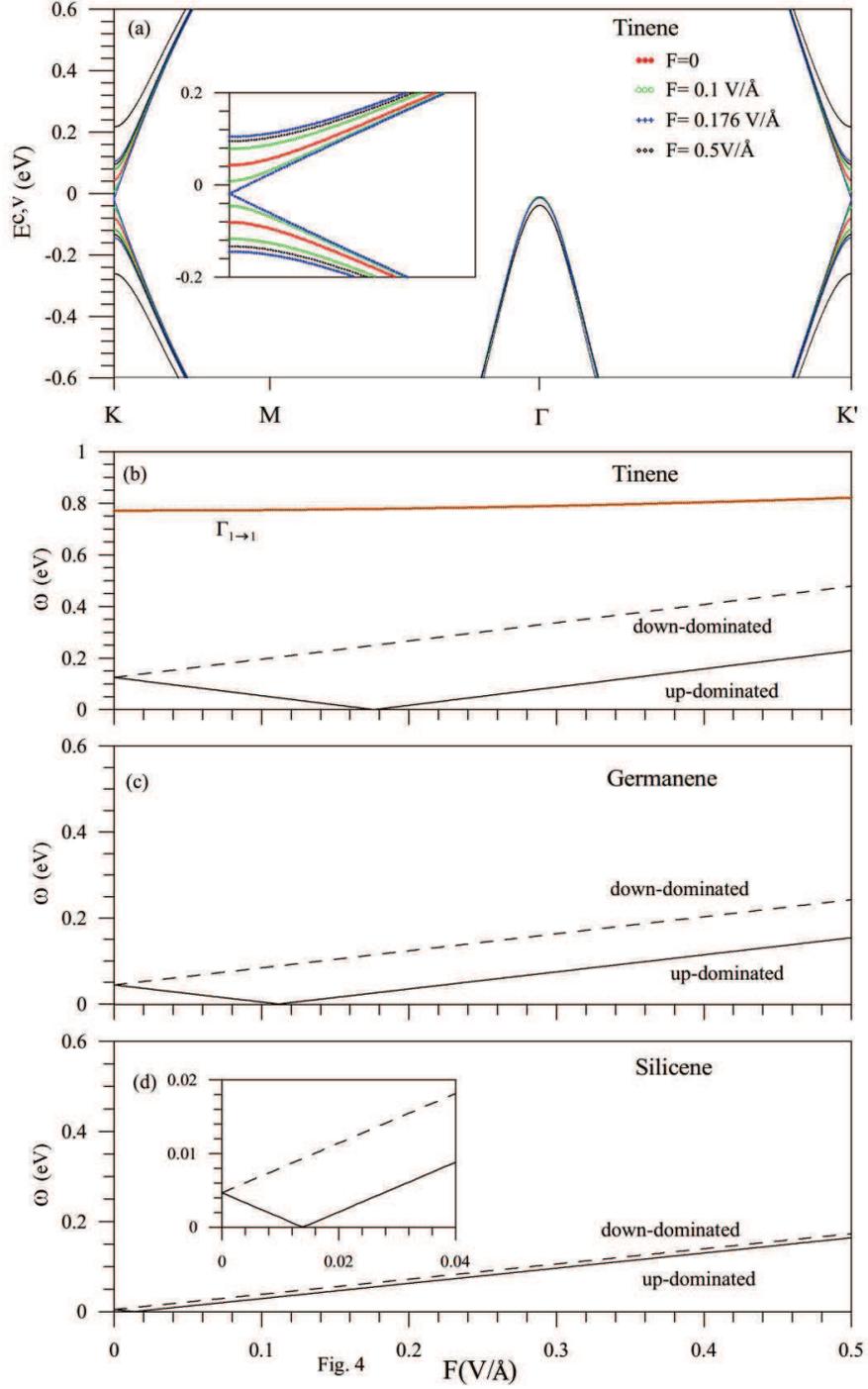}
\caption{(a)  Energy bands of tinene in the range of $-$0.6 eV${\le\,E^{c,v}\le\,}$0.6 eV at various electric fields, and (b) the $F-$dependent absorption frequencies within ${\omega\le\,1}$ eV for (b) tinene, (c) germanene and (d) silicene.}
\label{FIG:4}
\end{figure}

\begin{figure}
\centering
\includegraphics[width=0.90\textwidth]{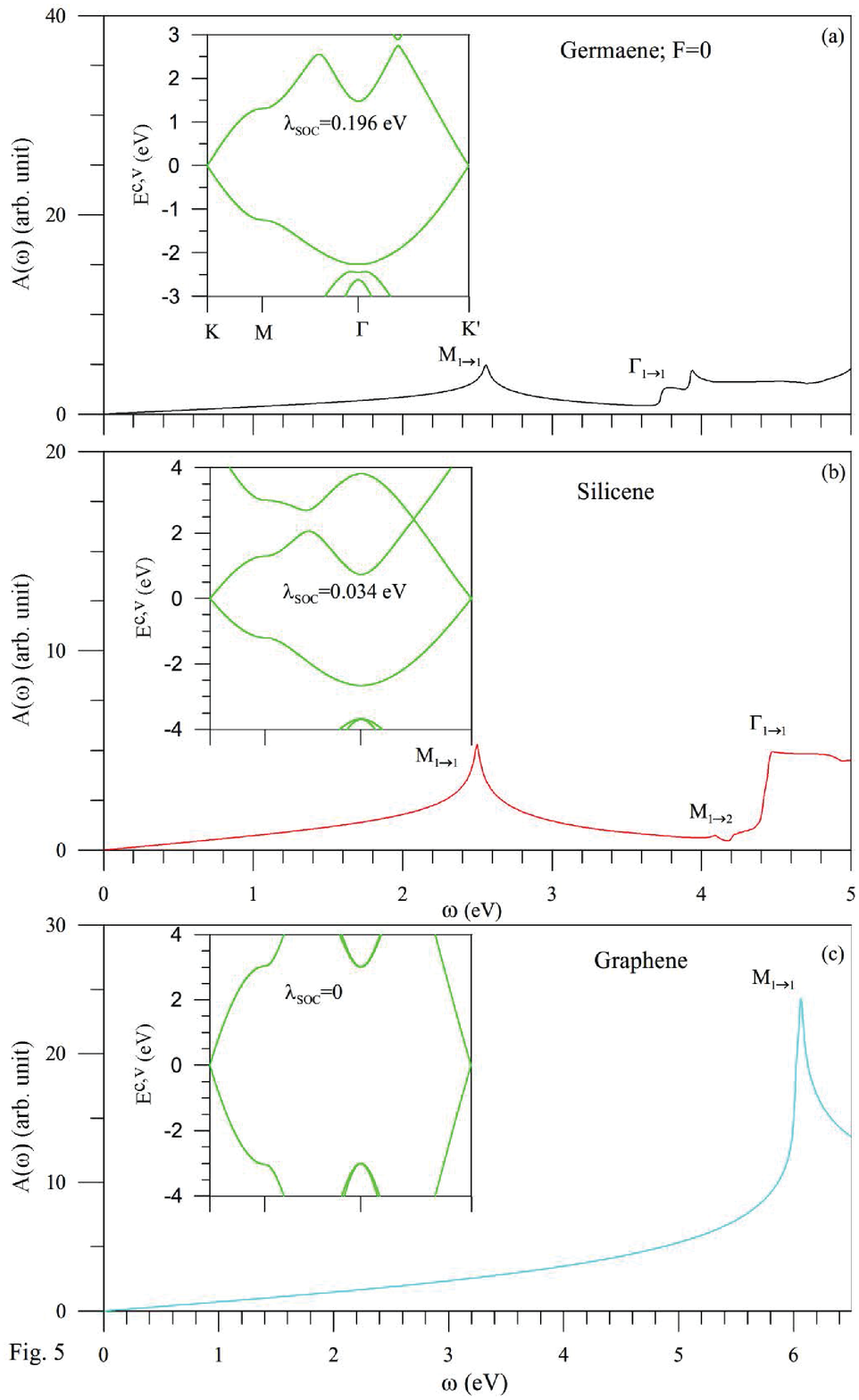}
\caption{Band structures and absorption spectra for (a) germaene, (b) silicene, and (c) graphene.}
\label{FIG:5}
\end{figure}

\end{document}